\documentclass[11pt,a4paper]{article}

\usepackage{jcappub}
\usepackage{epstopdf}
\DeclareGraphicsExtensions{.pdf,.png,.jpg}
\usepackage{amsmath}
\usepackage{amsfonts}
\usepackage{amssymb}
\usepackage{url}

\title{Constraints on dissipative unified dark matter}

\author[a,b]{ Hermano~Velten}
\author[b]{and Dominik J. Schwarz}

\affiliation[a]{Universidade Federal do Esp\'irito Santo, Av. Fernando Ferrari, Goiabeiras, Vit\'oria, Brasil}

\affiliation[b]{Fakult\"at f\"ur Physik, Universit\"at Bielefeld, Postfach 100131, 33501 Bielefeld, Germany }

\emailAdd{velten@physik.uni-bielefeld.de}
\emailAdd{dschwarz@physik.uni-bielefeld.de}

\abstract{Modern cosmology suggests that the Universe contains two dark 
components -- dark matter and dark energy -- both unkown in laboratory 
physics and both lacking direct evidence. Alternatively, a unified dark 
sector, described by a single fluid, has been proposed. 
Dissipation is a common phenomenon in nature 
and it thus seems natural to consider models dominated by a viscous 
dark fluid. We focus on the study of bulk viscosity, as isotropy and 
homogeneity at large scales implies the suppression of shear 
viscosity, heat flow and diffusion. The generic ansatz $\xi \propto \rho^{\nu}$ for the coefficient of bulk viscosity ($\rho$ denotes the mass/energy density), 
which for $\nu = - 1/2$ mimics the $\Lambda$CDM background evolution, 
offers excellent fits to supernova and H(z) data.  
We show that viscous dark fluids suffer from large contributions 
to the integrated Sachs-Wolfe effect (generalising a previous study by 
Li $\&$ Barrow) and a suppression of structure growth at small-scales 
(as seen from a generalized Meszaros equation). 
Based on recent observations, we conclude that viscous dark fluid models 
(with $\xi \propto \rho^{\nu}$ and neglecting baryons) are strongly challenged. 
}


\begin{document}
\maketitle


\section{Introduction}

The cosmological concordance model states that the Universe is spatially flat and
approximately 
$95 \%$ of its energy content is made up of an unknown dark sector. The remaining
$5\%$ is known: baryonic matter, electrons, photons and neutrinos. In the context of the concordance
model a 
quarter of the dark sector behaves like cold dark matter (CDM), a pressureless component
that clusters. 
The remaining dark stuff is called dark energy and is hold responsable for the
current accelerated expansion of the Universe.

Dark energy is commonly modeled either by a scalar field or a dissipationless fluid. 
The isotropy and homogeneity of the Universe at large scales, suggests that such a
fluid description 
makes sense, at least as an effective description. Then dark energy has negative
pressure, such that 
its equation of state $p \sim - \rho$ today. Alternatively, one might view dark
matter and dark energy as different manifestations of one single substance  --
unified dark matter. Its equation of state function
$w \equiv p/\rho$ must be time-dependent in order to interpolates from the matter
dominated epoch 
to the current accelerated expansion. 

In this work we study a single-fluid descriptions of the dark
sector. As long as only the homogeneous and isotropic
evolution (the background) is concerned, 
this class of models is indistinguishable from the concordance model. However, when
it comes to perturbations differences become apparent at late times. We investigate in detail their contributions to the integrated Sachs-Wolfe effect, which
probes most efficiently the large scales at 
late times, and the small-scale matter power spectrum, which probes the other side
of the structure formation. We demonstrate that a generic single-fluid description of the
Universe seems to be strongly challenged from both sides.   

A prominent candidate for this scenario is the Chaplygin gas, where $p_{\rm
c}=-A/\rho_{\rm c}$, 
and its generalized version $p_{\rm gc}=-A/\rho^{\alpha}_{\rm gc}$ \cite{GCG}. For
the case 
$\alpha = 1$, this exotic equation of state is motivated by string theory, where the 
Chaplygin gas is interpreted as an effective description of a gas of D-branes in a 
D+2-dimensional space-time \cite{jackiw}. For the background evolution, 
$\alpha = 0$ corresponds to the $\Lambda$ cold dark matter model ($\Lambda$CDM).  

Another popular candidate is a dissipative fluid with intrinsic bulk viscosity
\cite{viscous,ViscousWS}. 
Any real fluid shows dissipative phenomena and thus it is well motivated to include
this aspect in 
cosmology as well. Typically  shear viscosity is more important than bulk 
viscosity, however isotropy and homogeneity of the Universe at large scales, does
not allow for 
shear. From the same argument diffusion and heat conduction cannot play an important
role. 
Thus at large scales, bulk viscosity must be the dominant dissipative effect.   
For an expanding universe it gives rise to a negative contribution to pressure, 
$p_{\rm v}=- \xi \Theta$, where $\Theta$ is the volume expansion rate of the fluid
and $\xi > 0$ is the 
coefficient of bulk viscosity. In order to study the cosmic expansion, one needs to
specify 
$\xi = \xi(t)$. A common ansatz, which we will adopt below, is $\xi \propto
\rho^{\nu}$ \cite{coeffxi}.

The generalized Chaplygin gas (GCG) and the viscous dark fluid (VDF) have the same 
background dynamics in the one-fluid approximation \cite{velten}. However, the
difference 
between these fluids appears at perturbative level. While the GCG is viewed as a
dissipationless 
fluid, it has isentropic (the same entropy everywhere) perturbations 
$\delta p_{\rm gc}=(\dot{p}_{\rm gc}/\dot{\rho}_{\rm gc}) \delta\rho_{\rm gc}$, 
the perturbative dynamics of the VDF is, by definition, nonadiabatic and it could
also be 
non-isentropic, $\delta p_{\rm v} \neq (\dot{p}_{\rm v}/\dot{\rho}_{\rm v})
\delta\rho_{\rm v}$.  

As GCG and VDF are equivalent w.r.t. their background evolution, both models can fit
probes that 
are not sensitive to structure formation itself, e.g. SNIa data \cite{backresults}.
At the perturbative level one obtains, from the matter power spectrum and from the
cosmic microwave background (CMB) spectrum, very different predictions. The
confrontation of the GCG model with the matter power 
spectrum data, provided by the Sloan Digital Sky Survey (SDSS) and the Two Degree
Field Galaxy Redshift Survey (2dFGRS) data sets, discards a one-fluid GCG universe
\cite{waga} due to strong oscillations in its theoretical power spectrum. However,
it has been demonstrated that this problem is solved when a baryonic component is
taken into account \cite{baryons} or, if {\it ad hoc} entropy perturbations are
included in the GCG perturbative dynamics \cite{entropicGCG}. While the former is a
trivial solution, as the observed power spectrum corresponds to the visible matter and not
the dark one, the latter can be seen as a motivation to explore viscous
(intrinsically nonadiabatic) cosmologies. On the other hand, the bulk viscous model
does not show the same pathologies as the GCG, thanks to its nonadiabatic behavior
\cite{velten,velten2}.  The GCG with $\alpha \approx 0$ agrees with CMB data
\cite{carturan}, while, apparently, the VDF does not, due to a huge amplification of
the integrated 
Sachs-Wolfe (ISW) signal \cite{barrow}.

The authors of \cite{barrow} showed that, the evolution of the gravitational
potential in a VDF model differs from the $\Lambda$CDM model at late times, implying
a huge ISW effect. This is also found for the GCG \cite{carturan} and for general
unified dark matter cosmologies relying on a single 
scalar field \cite{bertacca}. However, these studies have been limited to fixed
values of the cosmological parameters and it remains unclear, if the huge ISW effect
could be avoided in a different region of parameter space. We study the dependence
of the ISW effect on the model parameters of 
VDF models and GCG models. 

We also study the behaviour of sub horizon pertubations in the VDF and GCG models 
during the matter dominated epoch. We show that structure formation can be
drastically affected 
in such cosmologies by comparing the growth of the unified dark matter perturbations with a typical CDM scenario. 
In other words, we investigate whether dark halos, the hosts of galaxies can form at
all. 

In the next section, we compare the background evolution of the GCG and VDF models
with 
$\Lambda$CDM. Section 3 is devoted to the study of linear perturbations. We provide
an evolution equation for the study of the ISW effect and we obtain Meszaros-like
equations for the evolution 
of sub-horizon perturbations in the VDF and GCG models. In section 4 we derive
quantitative 
results for unified dark matter cosmologies and conclude with some remarks and open
issues 
in the final section. 
         
\section{Background evolution}

In this work, we assume a spatially flat one-fluid description of the matter content
of the Universe. 
This ansatz is expected to be appropriate at late times (thus radiation is
negligible). 
At small scales we also neglect the effects of baryonic matter, which limits the
precision of our 
discussion to the 10\% to 20\% level at small scales. 

The description of relativistic viscous fluids allows for a freedom in the choice of
the comoving frame. Comoving observers could be comoving with energy transport
(Landau-frame) or with particle number transport (Eckart-frame). Both approaches are
equivalent, but one has to make a choice. Here we adopt the Eckart formalism
\cite{eckart}. Then, the VDF bulk pressure is given by $p_{\rm v} = -\xi \Theta$. 
Due to the second law of thermodynamics the coefficient of bulk viscosity $\xi \geq 0$.
The volume expansion rate $\Theta \equiv u^{\mu}_{;\mu}$ (Greek indices run from 0
to 3, ";" denotes a covariant derivative) is obtained from the fluid velocity
$u^\mu$.  In a homogeneous and isotropic Universe, $\Theta = 3 H$, where $H$ is the
Hubble expansion rate. With the ansatz 
\begin{equation} 
\xi=\xi_0\left(\frac{\rho}{\rho_{0}}\right)^{\nu}, 
\end{equation} 
and assuming that the kinetic pressure $p=0$, the bulk viscous pressure of the 
background becomes [by means of $H = H_0(\rho/\rho_0)^{1/2}$]
\begin{equation}
p_{\rm v} = - 3H_0\xi_0\left(\frac{\rho}{\rho_{0}}\right)^{\nu+1/2}. 
\end{equation}

The GCG model has a similar equation of state, $p_{\rm gc} = - A
\rho_{0}\left(\rho_{0}/\rho\right)^{\alpha}$, with $A$ and $\alpha$ being 
dimensionless parameters. 

In a perfectly homogeneous and isotropic Universe, the GCG model and the 
VDF are equivalent, which is easily verified by the replacements
$\alpha=-(\nu+\frac{1}{2})$ and 
$A = 3 H_0 \xi_0/\rho_0$. Hence, both fluids show the same time evolution. Instead
of $\xi_0$ or 
$A$, it also is convenient to use the deceleration parameter $q_0$. This
correspondence can be established by 
\begin{equation} 
q_0=\frac 12 (1 - 3 A) = \frac{1}{2}\left(1- \frac{9H_0\xi_0}{\rho_0}\right).
\end{equation}
Once $\xi_0>0$, then $q_0 < 1/2$. The background evolution of the VDF and the GCG is
governed 
by ($a$ denotes the scale factor and $a_0 = 1$)
\begin{equation}
\left(\frac{H_{\rm v}}{H_0}\right)^{2}=\left[\frac{3 H_0 \xi_0}{\rho_0} + 
\frac{1 - \frac{3 H_0
\xi_0}{\rho_0}}{a^{3\left(\frac{1}{2}-\nu\right)}}\right]^{\frac{1}{\frac{1}{2}-\nu}}
\end{equation} 
and
\begin{equation}
\left(\frac{H_{\rm
gc}}{H_0}\right)^{2}=\left[A+\frac{1-A}{a^{3\left(1+\alpha\right)}}\right]^{\frac{1}{1+\alpha}},
\label{rhoH}
\end{equation}
respectively. The existence of an early matter dominated epoch, $H(a \ll 1) \sim
a^{-3/2}$, is guaranteed 
for $\nu<1/2$ and $\xi_0 < \rho_0/(3H_0)$ for the VDF model and for 
$\alpha>-1$ and $A < 1$ in the GCG case. In order to obtain an accelerated epoch at
late times 
($q_0 < 0$), the parameters must obey $\xi_0 > \rho_0/(9 H_0)$ and $A > 1/3$,
respectively. 
The early and late time limits of both models are equivalent to the $\Lambda$CDM
model. 
The only difference is the transition from the matter dominated phase to the
accelerated epoch, 
which is given by the equation of state functions
\begin{equation}
w_{\rm v} \equiv \frac{-3H\xi}{\rho} 
= \frac{-1}{1+\frac{\rho_0 - 3 H_0 \xi _0}{3 H_0 \xi_0}(1+z)^{3(\frac{1}{2}-\nu)}}
\end{equation}
and
\begin{equation} 
w_{\rm gc}=\frac{p_{\rm gc}}{\rho}=\frac{-1}{1+\frac{(1-A)}{A}(1+z)^{3(1+\alpha)}}.
\end{equation}
The expressions in (\ref{rhoH}) are analogue to the
$\Lambda$CDM one,
\begin{eqnarray}
\left(\frac{H_{\Lambda}}{H_0}\right)^{2}=\frac{\Omega_{m0}}{a^{3}}+1-\Omega_{m0},
\end{eqnarray}
if we adopt $q_0=\frac{3\Omega_{m0}}{2}-1$ $(A=1-\Omega_{m0})$ and $\nu=-1/2$
$(\alpha=0)$ for the VDF (GCG) model. These relations will be usefull in the next
section in order to compare the perturbative dynamics of these models. 

\section{Density perturbations of a dissipative fluid}

In this section we study the perturbative dynamics for the VDF and the GCG models.
The differences 
between both models for an inhomogeneous Universe can be traced back to an inherent 
nonadiabatic behavior of the viscous model. In a sense, the VDF model can be seen as a 
nonadiabatic version of the GCG model. 

Let us start by considering the most general dissipative fluid with energy momentum
tensor 
$T^\mu_\nu$, including a dissipative contribution which is denoted by $\Delta
T^{\mu}_{\,\,\nu}$. 
In the Eckart frame, the most general dissipative tensor is 
\begin{equation}
\Delta T^{\mu}_{\,\,\nu}=-\xi\Delta T^{\mu}_{b\,\,\nu}-\eta\Delta T^{\mu}_{s\,\,\nu}-
\kappa\Delta T^{\mu}_{h\,\,\nu},
\end{equation}  
where $\xi, \eta$ and $\kappa$ are the coefficients of bulk viscosity, shear
viscosity and heat conduction. For the homogeneous and isotropic background, only
the bulk viscosity contributes to the cosmic dynamics. At first order, the heat
conduction contributes only to the non-diagonal elements of 
$\Delta T^{\mu}_{\,\,\nu}$, and thus producing negligible contributions on
superhorizon scales. 
The same happens with shear viscosity. In contrast to bulk viscosity, shear
viscosity and heat conduction, influence the evolution of cosmological perturbations
via spatial gradients. 

In the following we neglect heat conduction and shear viscosity also at the level of
perturbations and 
thus the cosmic fluid is described by the energy-momentum tensor
\begin{eqnarray}
T^{\mu}_{\,\,\nu} &=& \rho u^{\mu}u_{\nu} + p h^{\mu}_{\nu} + \Delta
T^{\mu}_{\,\,\nu} =
\rho u^{\mu}u_{\nu} + p h^{\mu}_{\nu}
- \xi u^{\gamma}_{\,\,\,; \gamma} h^{\mu}_{\,\,\nu},
\label{EMT}
\end{eqnarray}
where $h^{\mu\,\nu}=g^{\mu\,\nu}+u^{\mu}u^{\nu}$.
More explicity, the background components of (\ref{EMT}) are
\begin{eqnarray}
T^{0}_{\,\,0}=-\rho, \quad T^{0}_{i}=T^{i}_{0}=0, \quad
T^{i}_{j} = p_{\rm eff} \delta^{i}_{j} = \left(p-\frac{3 \xi
\mathcal{H}}{a}\right)\delta^{i}_{j},
\end{eqnarray}
where $\mathcal{H}=\frac{a^{\prime}}{a}$ and the symbol ($'$) means derivative wrt
the conformal 
time $\eta$. 
Latin indices run from 1 to 3. The effective pressure $p_{\rm eff}$ is the sum of an
adiabatic 
component and the bulk viscous pressure (nonadiabatic). The VDF model is specified
by $p=0$ and 
a dissipationless fluid is recovered with $\xi=0$.

In the conformal Newtonian gauge the line element for scalar perturbations of an
isotropic and homogeneous, spatially flat universe is
\begin{eqnarray}
ds^{2}=a^{2}\left(\eta\right)\left[-\left(1+2\phi\right)d\eta^{2}+\left(1-2\psi\right)\delta_{ij}dx^{i}dx^{j}\right].
\end{eqnarray}
The linear perturbations of the fluid 4-velocity are given by 
\begin{eqnarray}
u^{0}=\frac{1}{a}(1-\phi), \quad u_{0}=-a(1+\phi),  \quad
u^{\gamma}_{;\gamma}=\frac{3\mathcal{H}}{a}+\delta u^{i}_{,i}-
\frac{3\mathcal{H}\phi}{a}-\frac{3\psi^{\prime}}{a}.
\end{eqnarray}

For the linear perturbations of (\ref{EMT}) we define the velocity scalar $v$, which
is associated with 
the peculiar velocity by $\delta u^{i}_{,i} \equiv - kv/a$, where $k$ is the
comoving wavenumber. 
The perturbed components of (\ref{EMT}) read
\begin{eqnarray}
\delta T^{0}_{\,\,0} &=& -\delta\rho, \\
\delta T^{0}_{i} &=& \frac{\rho}{a}(1+w+w_{\rm v})\delta u_{i}, \\
\delta T^{i}_{j} &=& \delta p\delta^{i}_{j}+\left[\xi(\frac{k
v}{a}+\frac{3\mathcal{H} \phi}{a}+\frac{3\psi^{\prime}}{a})-\frac{3\mathcal{H}}{a}
\delta\xi \right]\delta^{i}_{j}.
\end{eqnarray}
$\delta \xi$ denotes the perturbation of the coefficient of bulk viscosity. 
The adiabatic speed of sound $c^{2}_S \equiv (\partial p/\partial \rho)_S$. For 
dissipationless fluids, $c^2_S = p'/\rho'$ for the purposes of linear perturbation
theory. For dissipative 
fluids in linear perturbation theory $c^2_S = (p'/\rho')_{\xi = 0}$.

As we neglect anisotropic stresses in our model, the spatial off-diagonal Einstein 
equation implies $\phi=\psi$. At first order, the (0-0), (0-$i$) and the ($i$-$i$)
components of the 
perturbed Einstein equation read ($\Delta \equiv 
\delta\rho/\rho$)
\begin{eqnarray} 
-k^{2}\psi-3\mathcal{H}\psi^{\prime}-3\mathcal{H}^{2}\psi = \frac 32 \mathcal{H}^2
\Delta,
\label{poisson}
\end{eqnarray}
\begin{eqnarray}
-k\left(\psi^{\prime}+\mathcal{H}\psi\right)= \frac 32 (1 + w + w_{\rm v})
\mathcal{H}^2 v,
\label{Einstein2}
\end{eqnarray}
\begin{eqnarray}
\psi^{\prime\prime} + 
3\mathcal{H}\psi^{\prime} - 
(w + w_{\rm v}) 3 \mathcal{H}^{2} \psi
= \frac{3\mathcal{H}^{2}}{2}
\left[ \frac{\delta p}{\rho}-\frac{w_{\rm v}}{3\mathcal{H}}\left(kv
+3\mathcal{H}\psi+3\psi^{\prime}\right)+w_{\rm v}\frac{\delta\xi}{\xi}\right].
\label{Einstein3}
\end{eqnarray}

The pressure perturbation $\delta p = c_S^2 \delta \rho + \tau \delta S$, where
$\delta S$ denotes entropy perturbations and $\tau \equiv (\partial p/\partial S)_\rho$. 
Below we assume that pressure perturbations do not give rise to spatial fluctuations of the entropy to baryon ratio. 

\subsection{The integrated Sachs-Wolfe effect}

The ISW effect is a net change in the energy of a CMB photon as it passes through
evolving gravitational potential wells. It can be computed by
\begin{equation}
\left(\frac{\Delta T}{T}\right)_{\rm
ISW}=2\int^{\eta_{0}}_{\eta_{r}}d\eta\frac{\partial\psi}{\partial
\eta}\left[\left(\eta_{0}-\eta\right){\bf \hat{ n}},\eta\right],
\label{ISW}
\end{equation}
The integration is along the photon trajectory (${\rm \hat{n}}$) from $\eta_{\rm r}$ (conformal time
at recombination) to 
$\eta_0$ (conformal time today).

Combining the equations (\ref{poisson}) -- (\ref{Einstein3}) into a single
expression for the 
gravitational potential, we end up with
\begin{eqnarray}
\psi^{\prime\prime} + 
\left(1+c^{2}_{S}\right)3\mathcal{H} \psi^{\prime} + 
\left[\left(c^{2}_{S} - w \right) 3 \mathcal{H}^{2} +  c^{2}_{S} k^2\right] \psi = 
\qquad \qquad \qquad \qquad 
\nonumber \\
w_{\rm v}\left[
\left[-\frac{1}{2}+\frac{k^{2}}{(1+w+w_{\rm v})9\mathcal{H}^2}\right] 
3 \mathcal{H} \psi^{\prime} +
\left[\frac{3\mathcal{H}^{2}}{2}+\frac{k^{2}}{3(1+w+w_{\rm v})}\right] \psi +
\frac{3\mathcal{H}^2}{2}\Xi \right],
\label{evol1}
\end{eqnarray}
where $\Xi \equiv \delta \xi/\xi$ can be considered as the relative perturbation of the coefficient of bulk viscosity.

If we neglect the VDF contribution to the energy-momentum tensor, the right hand
side of equation (\ref{evol1}) vanishes and hence the resulting equation is the full
evolution for the gravitational potential of an adiabatic fluid with an equation of
state parameter $w=p/\rho$. The right hand side of (\ref{evol1}) represents the
influence of nonadiabaticity on $\psi$. For the VDF model, we set $c^{2}_S = w = 0$ and use the appropriate functions $w_{\rm v}$ and $H_{\rm v}$. For the last term we need to know the functional form of $\xi$. If $\xi = \xi_0 (\rho/\rho_0)^\nu$ its perturbation $\delta\xi = \nu \xi \Delta$ can be related to the potential $\psi$ using equation
(\ref{poisson}). 

\subsection{Evolution of sub-horizon perturbations}

In the radiation era pressure suppresses the growth of structures. However, cold
dark matter, once kinetically decoupled from the plasma, starts to grow
logarithmically on scales
smaller than the Hubble horizon even during this epoch. Once the
Universe becomes matter dominated ($z_{\rm eq} \sim 3000$) CDM can grow linearly in
the scale 
factor. This scenario is called hierarchical structure formation as smallest
structures form
first and later on merge and grow to evolve into larger structures. 

The unified dark matter models studied in this paper have a matter-like behavior in
the past, 
but do not necessarily provide a successful structure formation scenario. In order
to study scales which entered the horizon sufficiently long before matter-radiation
equality, we make use of the 
covariant conservation of the energy-momentum tensor ($T^{\mu}_{\nu;\,\mu}=0$). 
The first-order continuity equation reads
\begin{eqnarray}
\Delta^{\prime} - 
3\mathcal{H}\Delta\left(w-c^{2}_{S}+w_{\rm v}\right) - 
\left(1+w+2w_{\rm v}\right)\left(k v +3\psi^{\prime}\right)
 -3\mathcal{H}w_{\rm v}( \psi - \Xi)=0,
\label{cons1}
\end{eqnarray}
and the Euler equation is
\begin{eqnarray}
v^{\prime}+\left[\mathcal{H}\left(1-3c^{2}_{S}-3w_{\rm v}\right)+\frac{w_{\rm
v}^{\prime}}{1+w+w_{\rm v}}-\frac{w_{\rm v}k^{2}}{3\mathcal{H}\left(1+w+w_{\rm
v}\right)}\right] v -
\qquad \qquad  \nonumber\\
\frac{w_{\rm v} k}{\mathcal{H}\left(1+w+w_{\rm v}\right)}\psi^{\prime}
+\frac{k (1+w)}{1+w+w_{\rm v}} \psi
+\frac{w_{\rm v} k}{1+w+w_{\rm v}} \Xi +
\frac{k c^{2}_{S}}{1+w+w_{\rm v}} \Delta =0.
\label{cons2}
\end{eqnarray}
For the adiabatic case there are many studies about the evolution of sub-horizon
scales, even considering the possibility of energy other than matter or radiation
\cite{subhorizon} or modified theories of gravity \cite{sub2}. However, the
clustering properties of nonadiabatic CDM have not yet been considered in much
detail. 

For the VDF model $(w = c_S^2 = 0)$, we can simplify equations (\ref{cons1}) and
(\ref{cons2}) and 
take the subhorizon limit of the Poisson equation to obtain
\begin{eqnarray}
\Delta^{\prime}-3\mathcal{H}w_{\rm v}\Delta&=&(1+2w_{\rm v})k v - 3 \mathcal{H}
w_{\rm v} \Xi \\
v^{\prime}+\left[\mathcal{H}(1-3w_{\rm v})+\frac{w^{\prime}_{\rm v}}{1+w_{\rm
v}}-\frac{k^{2}w_{\rm v}}{3\mathcal{H}(1+w_{\rm v})}\right] v &=&
-\frac{k\psi}{1+w_{\rm v}}+\frac{kw_{\rm v}\psi^{\prime}}{\mathcal{H}(1+w_{\rm
v})}-\frac{k w_{\rm v}\Xi}{1+w_{\rm v}}\\
-k^{2}\psi&=& \frac 32 \mathcal{H}^{2} \Delta
\end{eqnarray}

It is covenient to combine these equations to a single second-order differential
equation for 
$\Delta$ and to use the scale factor $a$ instead of conformal time. Hence, we obtain a 
Meszaros-like equation:
\begin{equation}
a^{2}\frac{\,d^{2}\Delta}{da^{2}}+\left[\frac{a}{H}\frac{d
\,H}{da}+3+A(a)+B(a)k^{2}\right]a\frac{\,d\Delta}{da}+\left[+C(a)+D(a)k^{2}-\frac{3}{2}\right]\Delta=P(a),
\label{small}
\end{equation}
\begin{eqnarray}
A(a)=-6w_{\rm v}+\frac{a}{1+w_{\rm v}}\frac{dw_{\rm v}}{da}-\frac{2a}{1+2w_{\rm
v}}\frac{dw_{\rm v}}{da}+\frac{3w_{\rm v}}{2(1+w_{\rm v})}\nonumber
\end{eqnarray}
\begin{eqnarray}
B(a)=-\frac{w_{\rm v}}{3a^{2}H^{2}(1+w_{\rm v})}\nonumber
\end{eqnarray}
\begin{eqnarray}
C(a)=\frac{3w_{\rm v}}{2(1+w_{\rm v})}-3w_{\rm v}-9w^{2}_{\rm v}-\frac{3w^{2}_{\rm
v}}{1+w_{\rm v}}\left(1+\frac{a}{H}\frac{dH}{da}\right)-3a\left(\frac{1+2w_{\rm
v}}{1+w_{\rm v}}\right)\frac{dw_{\rm v}}{da}+\frac{6aw_{\rm v}}{1+2w_{\rm
v}}\frac{dw_{\rm v}}{da}\nonumber
\end{eqnarray}
\begin{eqnarray}
D(a)=\frac{w^{2}_{\rm v}}{a^{2}H^{2}(1+w_{\rm v})}\nonumber
\end{eqnarray}
\begin{eqnarray}
P(a)= - 3 w_{\rm v} a \frac{d\,\Xi}{da} + 
3 w_{\rm v} \Xi \left[-\frac{1}{2} + \frac{9w_{\rm v}}{2} +\frac{-1-4w_{\rm
v}+2w_{\rm v}^2}{w_{\rm v}(1+ w_{\rm v})(1+2w_{\rm v})} a \frac{d\,w_{\rm v}}{da} - 
\frac{ k^{2}(1-w_{\rm v})}{3H^{2}a^{2}(1+w_{\rm v})}\right]\nonumber
\end{eqnarray}

The function $P(a)$ contains all contributions from the perturbation of the
coefficient of 
bulk viscosity $\delta\xi$. In the limit $w_{\rm v}=0$ we
obtain the standard equation for CDM perturbations with the solution $\Delta_{\rm cdm} \propto
a$. 

The above equations are solved numerically. However, in order to obtain some analytic predictions for $\Delta$, note that in the sub-horizon limit $k \gg \mathcal{H}$ we find 
\begin{equation}
a^{2}\frac{\,d^{2}\Delta}{da^{2}}+ B(a)k^{2} a\frac{\,d\Delta}{da}+ D(a)k^{2} \Delta=
- 3 w_{\rm v} a \frac{d\,\Xi}{da} -
 w_{\rm v} \Xi \frac{ k^{2}(1-w_{\rm v})}{H^{2}a^{2}(1+w_{\rm v})}.
\end{equation}
If we also send $w_{\rm v}$ to $-1$, the $k^2$ terms dominate and the equation is dominated by the first derivative term, thus one can expect exponential
damping.

\section{Observational Constraints}

\subsection{Supernova and $H(z)$ data} 

We employ a statistical analysis using recent $H(z)$ \cite{Hz} and the SN Ia constitution \cite{hicken} data sets, in order to constrain the parameters of the background model. 

The confidence contours for a set of parameters $\left\{\bf p\right\}$ are
obtained from the probability distribution function (PDF)
\begin{eqnarray}
P\left(\bf p\right)=\mathcal{B}e^{-\frac{\chi^{2}(\bf p)}{2}}\nonumber,
\end{eqnarray} 
where $\mathcal{B}$ is a normalization constant. For a given sample, lets say SN, $\chi^{2}$ is defined by
\begin{eqnarray}
\chi^{2}_{SN}(\bf p)=\sum_{i}\frac{\left[\mu^{th}_{i}(\bf p)-\mu^{obs}_{i}(\bf
p)\right]^{2}}{\sigma^{2}_{i}}.
\end{eqnarray}
The quantities $\mu^{th}_{i}$ and $\mu^{obs}_{i}$ are the theoretical and the
observed values, of the distance moduli and $\sigma_i$ denotes their
error for each data point $i$. For the $H(z)$ sample we replace $\mu$ by $H$.
Hence, for the joint analysis we use $\chi^{2}=\chi^{2}_{SN}+\chi^{2}_H$.

\begin{figure}[!h]
\begin{center}
\includegraphics[width=0.48\textwidth]{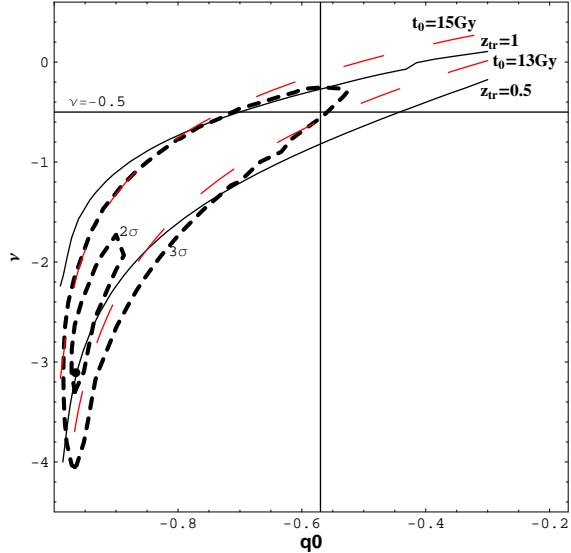}
\caption{Observational constraints from SN1a and H(z) data on the parameters 
of the VDF model ($q_0$ and $\nu$). The fat dot indicates the best-fit model. 
Short-dashed lines denote $2\sigma$ and $3\sigma$ contours. Long-dashed (red) 
lines indicate age constraints of 13 Gy and 15 Gy, respectively. 
Thin lines denote the redshift of the onset of accelerated cosmic expansion. 
The cosmic expansion history of a $\Lambda$CDM model is obtained 
for $\nu = -1/2$ (horizontal line) with its best fit at $q_0 = -0.57$ (vertical line).}
\label{background}
\end{center}
\end{figure}

Observational constraints on $q_0$ and $\nu$ are shown in Figure (\ref{background}). It displays the $2\sigma$ and $3\sigma$ confidence levels with best fit at $(q_0,\nu)=(-0.95,-3.2)$ with $\chi^{2}_{Viscous}=472.5$. The dashed-red lines are age constraints for which the Universe is $13$Gyr and $15$Gyr old. The parameters for which the transition to the accelerated epoch occurs at $z_{\rm tr}=1$ and $z_{\rm tr}=0.5$ are shown in the thin lines. We remark that these background results can be translated to the GCG model using the correspondences stablished in section 2. For the $\Lambda$CDM model (the horizontal line corresponding to $\nu=-0.5$) the best fit occurs at $q_0=-0.57$ (vertical line) that means $\Omega_{\Lambda}=0.71$. We obtain $\chi^{2}_{\Lambda CDM}=472.9$. 
Thus the latter $\chi^2$ is greater than for the viscous model. This occurs since the viscous model has an extra parameter. A model comparison by means of the Akaike information criterion,  $AIC=\chi^{2}+2k$ with $k$ being the number of free parameters \cite{AIC}, it becomes clear that both models are competitive with the $\Lambda$CDM model being slightly favoured ($|\Delta$AIC$| = 1.6$).  

\subsection{The integrated Sach-Wolfe effect from unified dark matter}

The CMB spectrum of anisotropies has been a key test for dark energy candidates as
well as for modified gravity theories. It has been observed that UDM models suffer
from an amplification of the ISW signal \cite{barrow}. In general, for the GCG, unless $\alpha=0$,
the acoustic peak to Sachs-Wolfe plateau ratio decreases for increasing $\alpha>0$. A similar conclusion
was obtained for the bulk viscous fluid in \cite{barrow}. However, the
dependence of these results on the free parameters of the UDM models is
still not clear and we adress this question now.  

We define a ''quality`` variable $Q_m$ to measure the difference between the ISW
signal for some model $m$ and the $\Lambda$CDM model,
\begin{equation}
Q_{m}\equiv\frac{\left(\frac{\Delta T}{T}\right)_{\rm ISW}^{m}}{\left(\frac{\Delta T}{T}\right)_{\rm ISW}^{\rm \Lambda CDM}}-1,
\label{Q}
\end{equation}
where positive (negative) values of $Q$ stand for an enhanced (a reduced) ISW effect for the
model $m$ to $\Lambda$CDM. The signal $\left(\frac{\Delta T}{T}\right)_{\rm ISW}^{m}$ can be obtained
from (\ref{ISW}), once we have calculated the gravitational potential $\Psi_m$ from
(\ref{evol1}). A similar
definition of $Q$ was considered in \cite{dent}. The relevant modes for the ISW
effect correspond to scales $k<0.003(h/{\rm Mpc})$, that is the approximate scale where
the Sachs-Wolfe $C_l$ plateau begins in the CMB temperature anisotropy angular power spectrum.
We shall plot the contours $Q=120\%,80\%,40\%$ and $0\%$ in parameter
space and compare them to the background constraints obtained above. With this
strategy we verify whether it is possible to conciliate the ISW effect contours close to $Q=0\%$ (i.e.~close to the $\Lambda$CDM model) with the ''allowed`` background model parameters. 

In order to estimate Q, we adopt a fiducial spatially flat $\Lambda$CDM model with parameters
$H_0=72$ km/s/Mpc and $\Omega_{m0}=0.266$, as suggested by WMAP-7.

One can ask if the current measurements of the ISW effect are able to discriminate between different models. In other words, are the values $Q=40\%$ and $80\%$ or even $Q=120\%$ acceptable? To answer this question we consider current and future estimations of the error bars of the CMB temperature and galaxy cross-correlation function $C_{gT}$, that is currently used to measure the ISW effect \cite{ISWmeasurement}. Since the ISW effect is hard to measure one can currently discard only the models with $Q>100\%$ (at 95\%C.L.), corresponding to $2\delta C_{gT}/C_{gT} \geq 1$. Radio surveys in the near future will reduce the error up to a factor of 5 and thus should be able to improve the limit to the $Q = 20\%$ level (see figure 9 of \cite{ISWfuture}).

\subsubsection{A model for $\xi(\rho)$ and its adiabatic counterpart}

With the ansatz $\xi(\rho)=\xi_0 (\rho/\rho_0)^{\nu}$ the quantity $\Xi$ becomes
\begin{equation}  
\Xi=\frac{2\nu}{3\mathcal{H}^2}\left(-k^{2}\psi-3\mathcal{H}\psi^{\prime}-3\mathcal{H}^{2}\psi\right).
\label{Xi}
\end{equation}

\begin{figure}[!h]
\begin{center}
\includegraphics[width=0.45\textwidth]{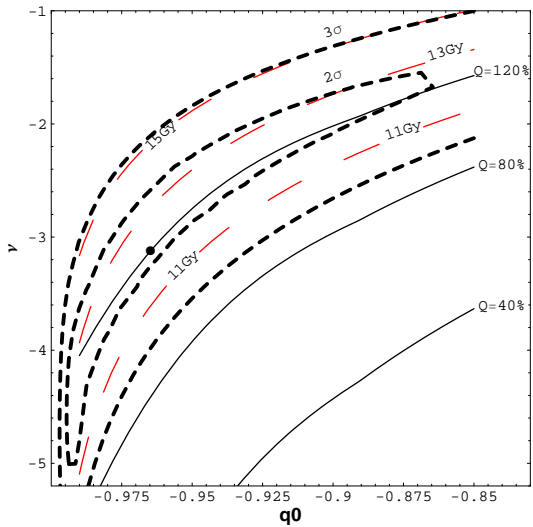}
\caption{Additional CMB temperature fluctuations from the integrated 
Sachs-Wolfe effect are estimated by $Q$, see (4.2). Solid lines represent 
contours of constant $Q_{\rm v}$ are shown in together with the $2\sigma$ and
$3\sigma$ contours and the age constraints of figure 1. VDF models with  
acceptable expansion history lead to at least a doubling of the ISW contributionwith respect to the WMAP 7yr best-fit $\Lambda$CDM model.} 
\label{fullbulk}
\end{center}
\end{figure}

The evolution equation for the gravitational potential is obtained by combining (\ref{Xi}) and (\ref{evol1}). For the bulk viscous model $(c^{2}_{s}=0)$ we solve it numerically and calculate $Q_{\rm v}$ (see (4.2)) for various choices of the background 
parameters. The results are shown in figure (\ref{fullbulk}). $Q=0\%$ and even $Q=40\%$ are in stark disagreement with the background
contours (short-dashed lines) which are compatible with the constraints obtained in
\cite{157Gold}. In additition, long-dashed lines display the age of the
universe with 11, 13 and 15 Gyrs. The best-fit model, symbol $\bullet$ in Figure
(\ref{fullbulk}), corresponds to $Q_{\rm v}=120\%.$ 

\begin{figure}[!h]
\begin{center}
\includegraphics[width=0.45\textwidth]{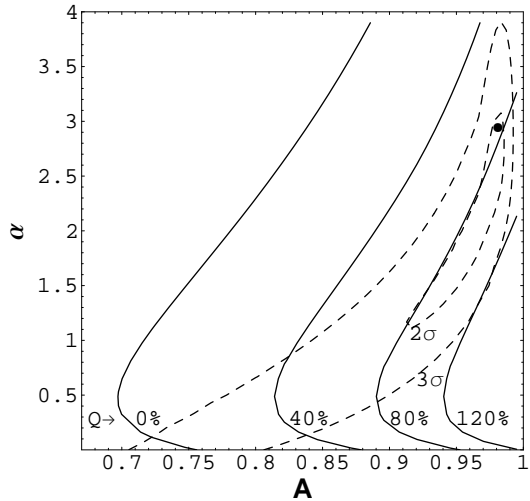}
\caption{As figure 3, but now for the GCG model. The free parameters of the GCG
model are $A$ and $\alpha$, as defined in the text. 
Dashed lines represent the $2\sigma$ and $3\sigma$ contours of the fit to SN1a
and $H(z)$ data, with best-fit denoted by the fat dot. 
From left to right, the solid lines are contours of constant ISW contribution
$Q_{\rm gc}=0\%,+40\%,+80\%$ and $+120\%$. For the GCG the ISW effect is 
slightly less pronounced compared to the VDF models.}
\label{cha}
\end{center}
\end{figure}

For the GCG model we compute equation (\ref{evol1}) with $w_{\rm v}=0$. Also we write
$\mathcal{H}$ as a function of $A$ and $\alpha$ and for the adiabatic speed of sound
we find
\begin{equation}
c^{2}_{\rm s\,gc}=-\alpha w_{\rm gc}=\frac{\alpha A}{A+(1-A)a^{-3(1+\alpha)}}.
\label{csch}
\end{equation}
We observe, see figure (\ref{cha}), a small improvement, as the best fit model is
close to the $Q_{\rm gc}=80\%$ line. However, both cases are discarded by the
CMB data and this result agrees with \cite{carturan,barrow,bertacca}.

\begin{figure}[!h]
\begin{center}
\includegraphics[width=0.45\textwidth]{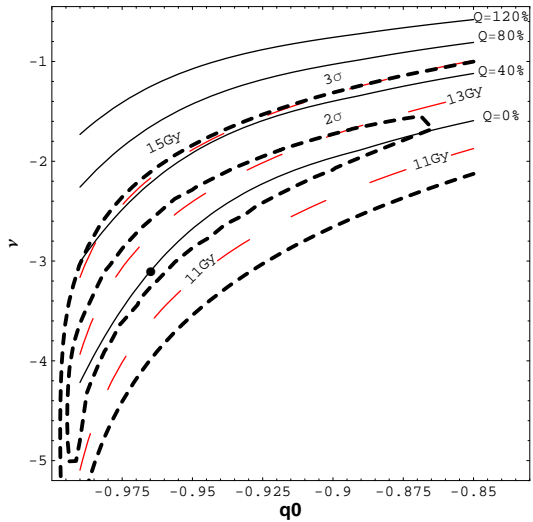}
\caption{As figure 2, but now the perturbation of the coefficient of bulk 
viscosity is arbitrarily set to zero, $\delta\xi = 0$. 
>From top to botton the solid lines are the corresponding 
$Q_{\rm v}=+120\%,+80\%,+40\%$ and $0\%$ contours.
Long-dashed (red) lines display the age of the universe with 11, 13 
and 15 Gy. Now, VDF models fit the background and do not show an enhanced
ISW effect. However, there is no physical motivation to put $\delta\xi = 0$ 
in the context of VDF models, unless $\nu = 0$. }
\label{bulkdeltaxi=0}
\end{center}
\end{figure}

In section 3 the perturbation of the bulk viscous coefficient
$(\delta \xi)$ has been considered as a free function with its effects gathered by $\Xi$.
Of course, a full perturbative analysis of the bulk viscous fluid should include
this term. Let us for a moment take the freedom to treat $\Xi$ as a free function of time
neglecting the form imposed by (\ref{Xi}). For the case $\Xi=cte=0$ we observe that
it is possible to conciliate the background constraints with the non-amplified ISW
effect line $(Q_{\rm v}=0)$ as shown in figure \ref{bulkdeltaxi=0}. We remark that
the background dynamics is exactly the same as before and the nonadiabatic
contributions, except for the term $\Xi$, are still active on the r.h.s.~of
(\ref{evol1}). The analysis shown in figure \ref{bulkdeltaxi=0} reveals that $\Xi \neq0$ is the source of 
the amplified ISW effect which has plagued the VDF model. For $\zeta \propto \rho^\nu$, we cannot regard 
$\Xi=0$ as a solution to the problem, since $\Xi = 0$ occurs only if $\nu=0$ or $\Delta =0$. We know that there are 
density fluctuations in the Universe, thus $\Delta \neq 0$. On the other hand, assuming $\nu=0$ leads to very different 
background dynamics, which we study below as a particular configuration of the VDF model.

We have also verified that extremely large negative (positive) values for
$\nu(\alpha)$ do not produce a large amplification in the ISW effect. Concerning the
GCG, this limit of the parameter $\alpha$ had been found before in \cite{oliver} but
the correspondence with $\nu$ had not yet been established. Large values
for the parameter $\nu$ also agree with the analysis using the matter power spectrum
\cite{velten}. This range for the parameter $\nu(\alpha)$ implies a
step-transition of the background evolution from a CDM phase to a deSitter one as
discussed in \cite{piattella2}.

\subsubsection{A constant coefficient of bulk viscosity}

\begin{figure}[!h]
\begin{center}
\includegraphics[width=0.5\textwidth]{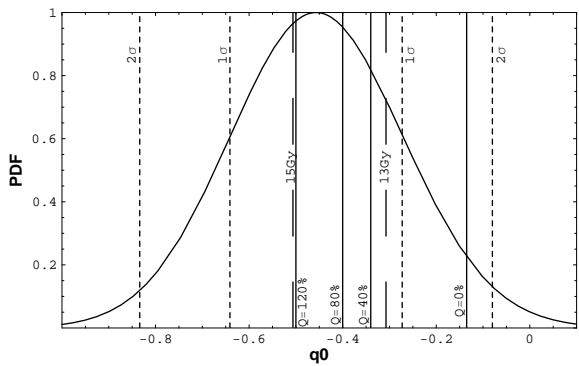}
\caption{PDF for the case $\nu=0$ with best fit at $q_0=-0.46$. The short-dashed
lines denote the $1\sigma$ and $2\sigma$ regions. The age constraints 
($13Gy$ and $15Gy$) are shown by long-dashed lines. Solid lines represent, 
from left to right, $Q_{\rm v}=120\%, 80\%, 40\%$ and $0\%$. The regions of 
a small enhancement of the ISW effect is in conflict with the age of the 
Universe.}
\label{nu=0}
\end{center}
\end{figure}

The previous considerations suggest to study the case of a constant coefficient of bulk viscosity ($\nu =0$) in more detail. Now, the VDF has only $q_0$ as free
parameter. At perturbative level there are no contributions from $\Xi$ but the rhs
of (\ref{evol1}) is non-vanishing. Figure \ref{nu=0} shows the PDF for $q_0$ with
the values $Q_{\rm v}=120\%,80\%,40\%,0\%$ and constraints from the age of the
universe. The line $Q_{\rm v}=0\%$ is within the $2\sigma$ region, but leads to a Universe younger than 13 Gyrs. In order to satistfy the age
constraints $Q>30\%$, which can be tested in the near future \cite{ISWfuture}.
 
\subsubsection{Mimicking the $\Lambda$CDM background evolution}

For $\nu=-0.5 (\alpha=0)$ the VDF (GCG) and the
$\Lambda$CDM models have exactly the same background evolution. Hence, the ISW contribution from nonadiabatic perturbations
can be quantified. Note that identical background solutions can be achieved by three different
models: i)the $\Lambda$CDM scenario, ii) a two-fluid model consisting of
pressureless, dissipationless matter and a dissipationless fluid with EoS $p=-\rho$, iii) the VDF(GCG) with
$\nu=-0.5(\alpha=0)$ which is also equivalent to a fluid with a negative constant
pressure. On the other hand, these models have distinct perturbative dynamics, namely: i) there are no perturbations from $\Lambda$, ii) could have nontrivial but adiabatic perturbations and iii) has nontrivial and non-adiabatic perturbations. The GCG is an example for ii). 

\begin{figure}[!h]
\begin{center}
\includegraphics[width=0.47\textwidth]{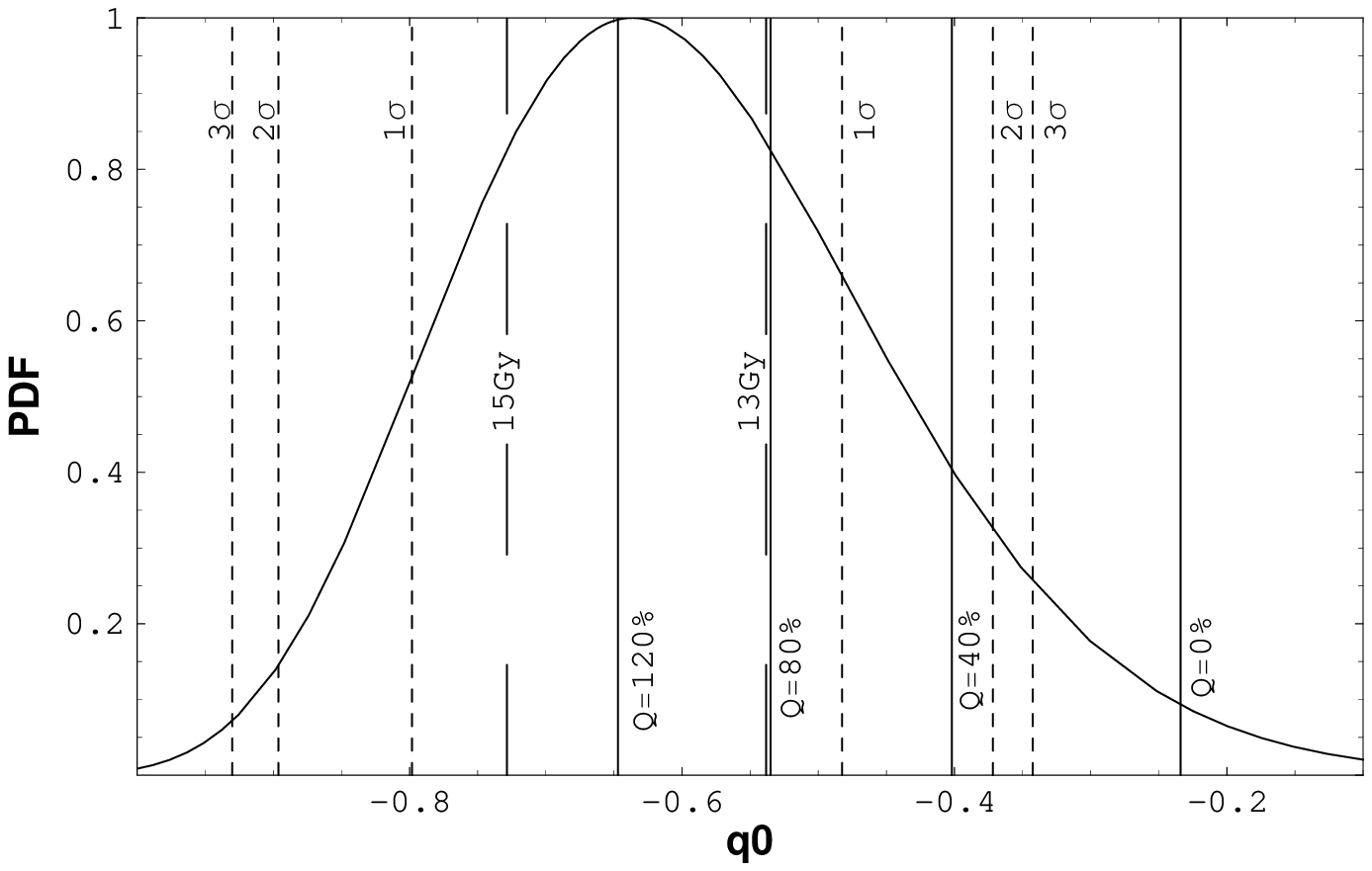}
\includegraphics[width=0.47\textwidth]{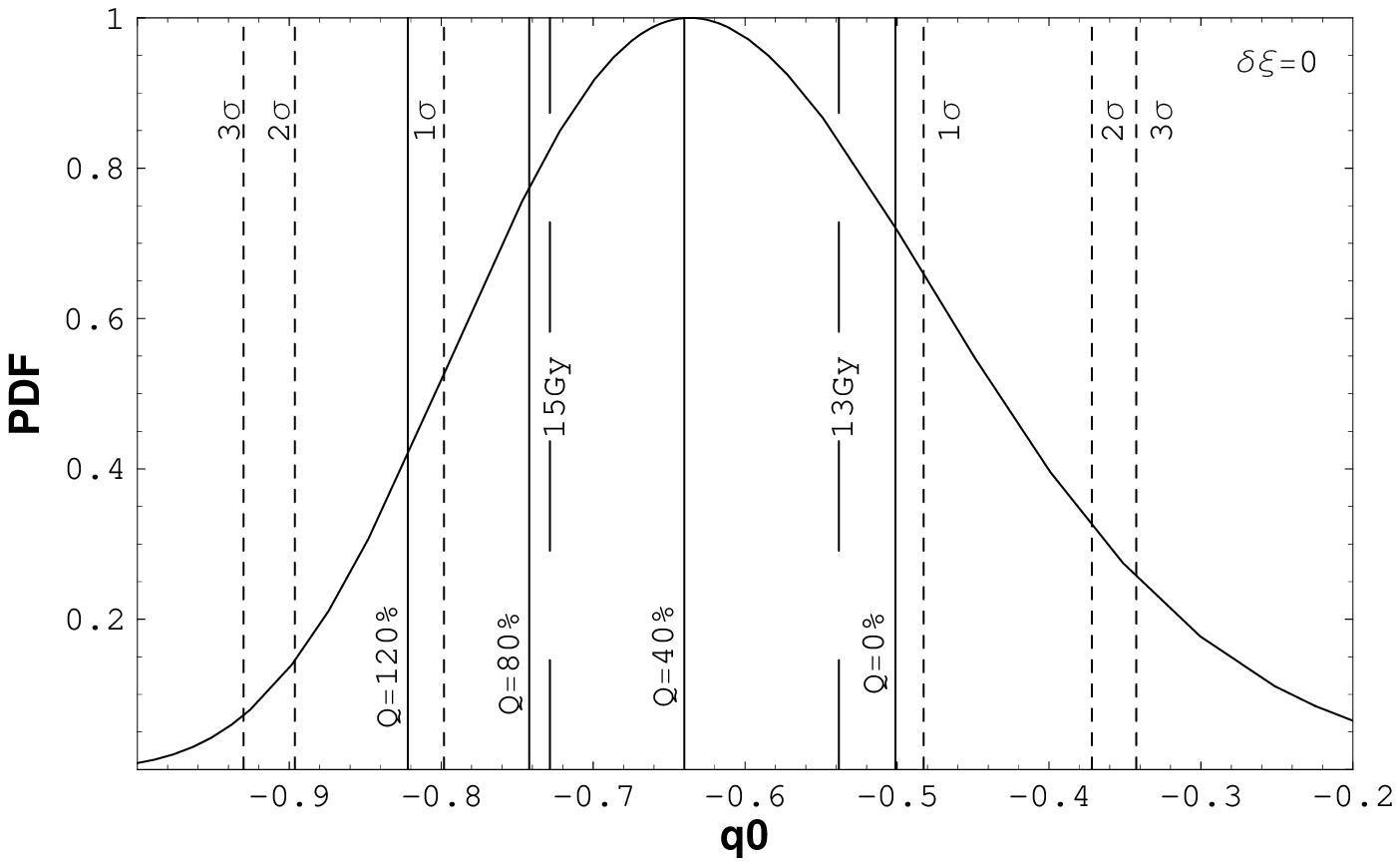}
\caption{PDFs for the VDF with $\nu=-0.5$. The best fit occurs at $q0=-0.64$. Left
panel shows the result considering the full evolution while in the right panel the
pertubation $\Xi$ was neglected. The age constraints ($13$ Gy and $15$ Gy) are
shown as long-dashed lines. Solid lines represent, from left to right,
$Q_{\rm v}=120\%, 80\%, 40\%$ and $0\%$.}
\label{nu-05}
\end{center}
\end{figure}

The result for the VDF with $\nu=-0.5$ is shown in the left panel in Figure \ref{nu-05}. The nonadiabatic contributions of the VDF are responsible for putting the $Q_{\rm v}=0\%$
line outside $3\sigma$ CL. However, if we neglect the
contribution from $\delta\xi$, the $Q_{\rm v}=0$ line is within the $1\sigma$ CL,
right panel in Figure \ref{nu-05}. 

\begin{figure}[!h]
\begin{center}
\includegraphics[width=0.47\textwidth]{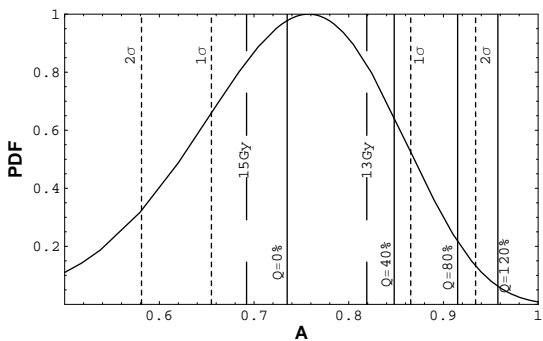}
\caption{PDF for the GCG with $\alpha=0$ and best fit at $A=0.76$. The age
constraints ($13$ Gy and $15$ Gy) are shown as long-dashed lines. Solid lines
represent, from left(right) to right(left), $Q_{\rm gc}=120\%, 80\%, 40\%$ and
$0\%$. This GCG model does not suffer from an ISW overproduction problem.}
\label{chaalpha0}
\end{center}
\end{figure}

Concerning the GCG with $\alpha=0$, the PDF for $A$ parameter is shown in Figure
\ref{chaalpha0}. This particular case behaves very similar to the $\Lambda$CDM and
our result agrees with \cite{carturan}.  

To summarize the observational constraints of sections 4.1 and 4.2, we have seen that generic VDF models that are excellent fits to SNIa and H(z) data sets, give rise to a large ISW contribution to the CMB temperature angular power spectrum and are thus excluded.

\subsection{Structure formation on small scales}

\begin{figure}[!h]
\begin{center}
\includegraphics[width=0.4\textwidth]{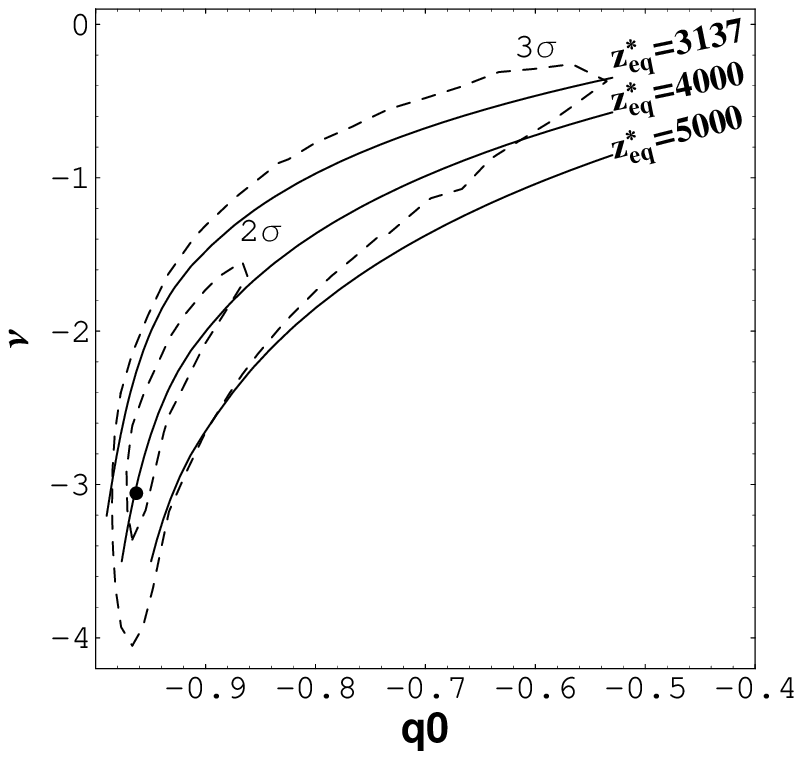}
\includegraphics[width=0.4\textwidth]{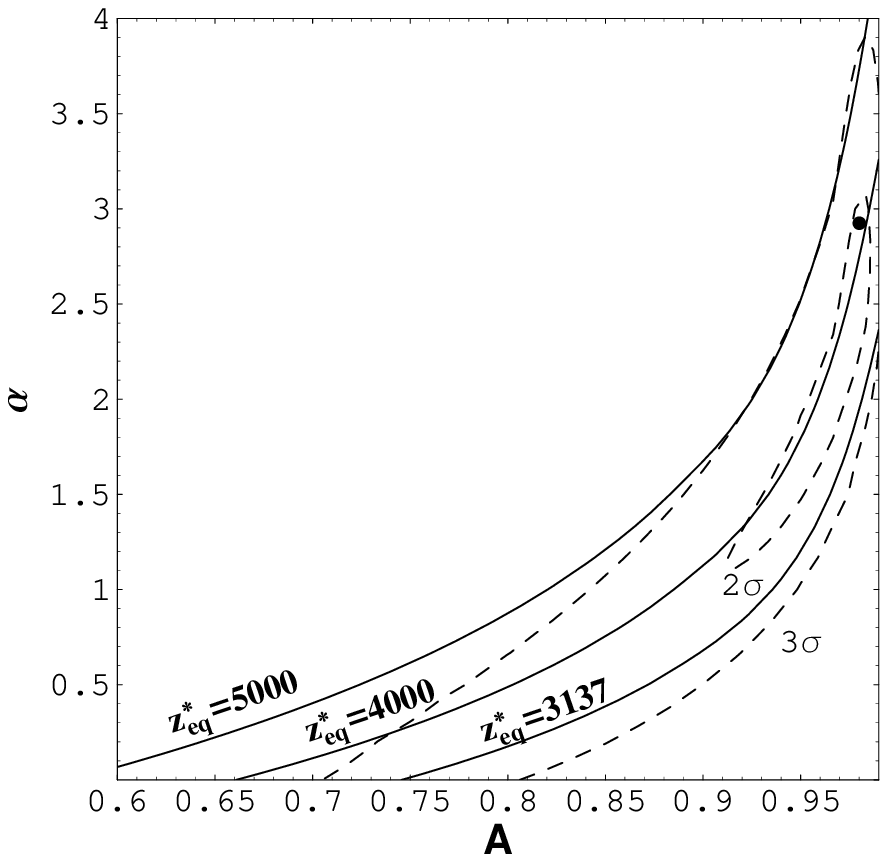}
\caption{Redshift of matter-radiation equality for the VDF (left) and 
GCG (right). Countours of constant $z_{\rm eq}^{*}$ (see text) are shown in 
the model parameter space of the VDF (left) and GCG (right). Dashed lines 
denote the constraints from SN and H(z) data as presented in figures 1 and 3, 
respectively.}
\label{Eq}
\end{center}
\end{figure}

In the standard CDM structure formation scenario small-scale perturbations 
start to grow $\propto a$ when the universe becomes matter dominated, at 
a redshift $z_{\rm eq}$. Before $z_{\rm eq}$, even if the wavelength of the 
perturbation is larger than the Jeans length rapid expansion prevents the 
growth of structures. Hence, before we can study the process of strucutre 
formation for the VDF it is essential to establish the time at which the
universe becomes VDF dominated and we thus include the radiation fluid
in our analysis. With the inclusion of radiation the dynamics of the 
GCG remains the same. However, since the expansion rate becomes
$H=[\frac{8\pi G}{3}(\rho_{\rm v}+\rho_r)]^{1/2}$ the background dynamics 
of the VDF is severely changed at early times. The fractional
density for the VDF is given by the numerical solution of
\begin{equation}
a\frac{d\Omega_{\rm v}}{da}+3\Omega_{\rm v}-\tilde{\xi}\Omega_{\rm
v}^{\nu}\left(\Omega_{\rm v}+\frac{\Omega_{\rm r0}}{a^{4}}\right)^{1/2}=0,
\label{Omegarv}
\end{equation}
where $\tilde{\xi}=9H_0\xi_0\rho_{c}^{\nu-1}$, $\rho_c$ is the critical density and
$\Omega_{\rm r0}=8.475\times10^{-5}$. The new model parameter $\tilde{\xi}$ 
is related to the deceleration parameter $q_0$ approximately by
\begin{eqnarray}
q_0=\frac{1}{2}\left(1+\frac{-9H_0\xi_0\rho_{\rm v0}^{\nu}+\rho_{\rm
r0}}{\rho_{\rm v0}+\rho_{r0}}\right)\approx\frac{1}{2}(1-\tilde{\xi}).
\label{xiq0}
\end{eqnarray}  
The fiducial $\Lambda$CDM model adopted in section 3 has the matter-radiation
equality occuring at $z_{\rm eq}=3137$. For a VDF plus
radiation the equality is a function of the model parameters and will be
denoted by $z_{\rm eq}^{*}$. As shown in figure \ref{Eq} for parameters 
values within $2\sigma$ CL, $z_{\rm eq}^{*}>z_{\rm eq}$. Hence, 
sub-horizon VDF fluid perturbations start to grow before typical CDM 
perturbations.

We solve equation (\ref{small}) with intial conditions
$\Delta_{\rm v}(z_{{\rm eq}^{*}})=1$ and
$\frac{d\Delta_{\rm v}}{da}(z_{{\rm eq}^{*}})=1$ and compare with the CDM
evolution $\Delta_{\rm cdm}\propto a$ with the same initial conditions, however
calculated at $z_{\rm eq}$, $\Delta_{\rm cdm}(z_{eq^{*}})=1$ and
$\frac{d\Delta_{\rm cdm}}{da}(z_{{\rm eq}^{*}})=1$. 
The Hubble rate and equation of state function  
in (\ref{small}) become
\begin{equation}
\left(\frac{H_{\rm v}}{H_0}\right)^{2}=\Omega_{\rm v}+\Omega_{\rm r0}a^{-4}
\hspace{1cm}w_{\rm v}=-\frac{1-2q_0}{3}(\Omega_{\rm v}+\Omega_{\rm
r0}a^{-4})^{1/2}\Omega_{\rm v}^{\nu},
\end{equation}
with $\Omega_{\rm v}$ being determined from (\ref{Omegarv}).

\begin{figure}[!h]
\begin{center}
\includegraphics[width=0.48\textwidth]{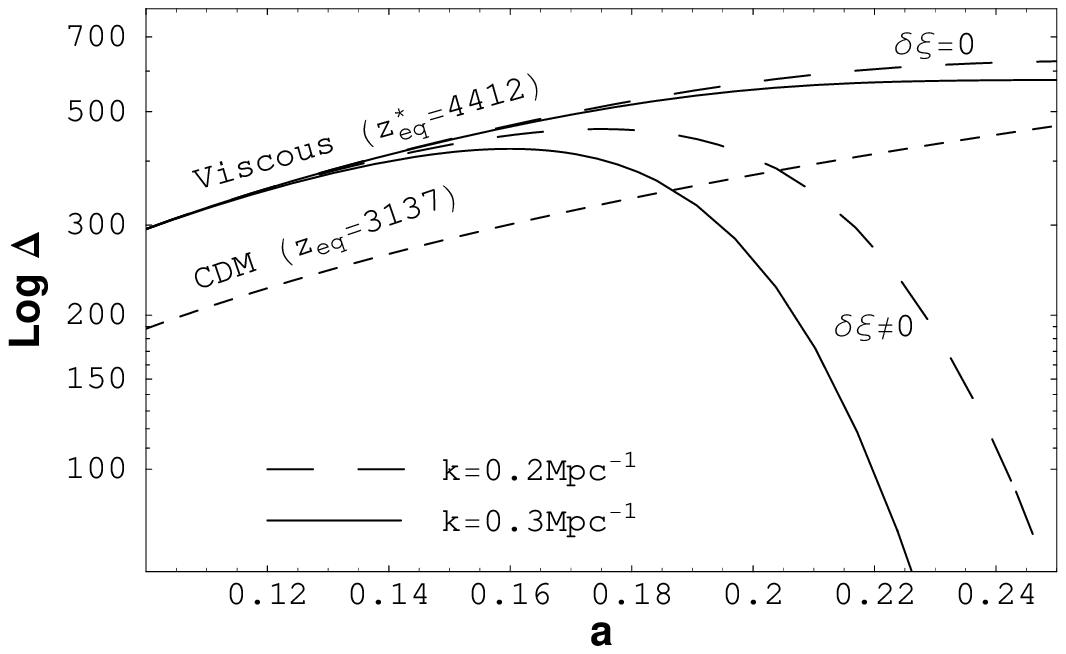}
\includegraphics[width=0.49\textwidth]{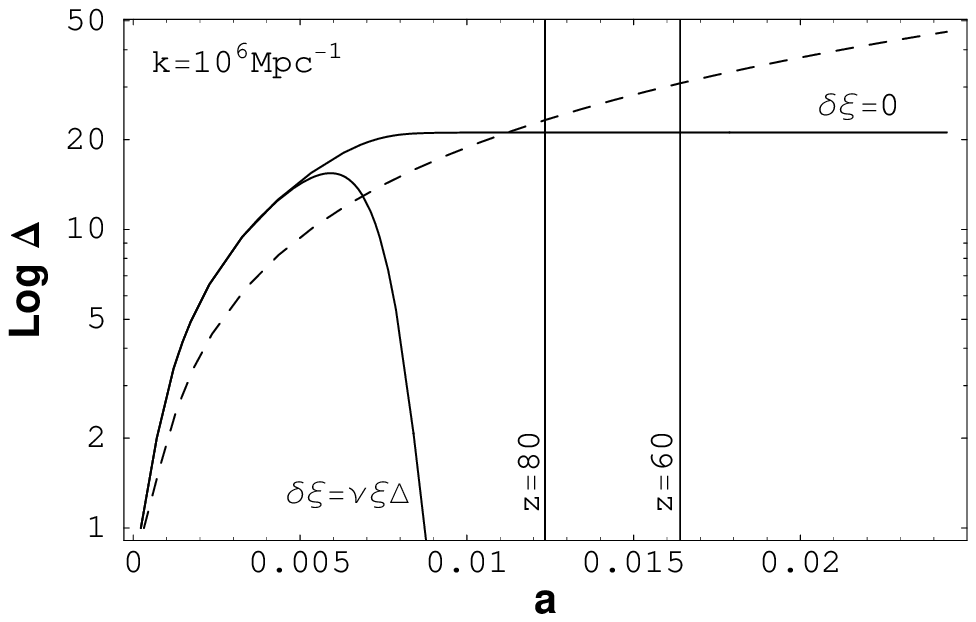}
\caption{Left panel: 
Growth of sub-horizon perturbations in CDM (short-dashed) in 
the $\Lambda$CDM model and of the VDF for $k=0.2 {\rm Mpc}^{-1}$ (long-dashed) 
and $k=0.3 {\rm Mpc}^{-1}$ (solid). The
upper lines for the viscous fluid have $\delta\xi=0$, while the bottom ones 
have $\delta\xi=\nu \xi \Delta$. 
Right panel: The same for $k=10^{6}{\rm Mpc}^{-1}$. Generic VDF models supress 
structures on subgalactic scales exponentially.}
\label{DeltaBestFit}
\end{center}
\end{figure}

We consider modes which give rise to cluster (subgalactic) size structures 
$k\sim 0.2{\rm Mpc}^{-1}$ ($k= 10^{6}{\rm Mpc}^{-1}$). We assume that for 
these modes the nonadiabatic Meszaros equation is valid up to the
onset of non-linear evolution at $z_{\rm nl}=3 (60\pm20)$ \cite{WimpyHalos}. 
Soon after $z_{\rm nl}$, a large fraction of the matter collapses into 
gravitationally bound objects. Nonlinear effects lead to a further modification
of the final (at $z=0$) power spectrum. The study of them is beyond the scope of
this work. Figure (\ref{DeltaBestFit}) shows the growth
of perturbations, considering the best fit model obtained above, for $k = 0.2$
-- $0.3
{\rm Mpc}^{-1}$ ($k=
10^{6}{\rm Mpc}^{-1})$ in left (right) panel. Also the CDM growth is shown as
short-dashed line. If we consider the full evolution of equation (\ref{small})
including the term $\Xi$ (botton lines indicated by $\delta\xi\neq0$) we observe a
large growth suppression after a redshift $z\sim6(a\sim0.14)$ for $k\sim
0.2{\rm Mpc}^{-1}$ and $z\sim200(a\sim0.005)$ for $k= 10^{6}{\rm Mpc}^{-1}$. Indeed, the
dominant contribution in the terms proportional to $k^{2}\Delta$ and
$k^{2}\Delta^{\prime}$ comes from $\Xi$ and, consequently, at late times the density
contrast $\Delta$ will decay rapidly. On the other hand, similarly to the ISW effect
results, the perturbative dynamics is well behaved if $\delta\xi=0$ (upper lines). 

The GCG perturbations do not suffer any kind of suppression and will
behave exactly like the CDM ones since it obeys the standard adiabatic growth
equation, $w_v=0$ in (\ref{small}), with solution $\Delta_{\rm gc}\propto a$. At the
same time, for any GCG configuration the transition to the accelerated
expansion phase occurs after $z_{nl}$ and the growth of perturbations is not
suppressed by this effect.      

\section{Conclusions}

The main idea behind the unification scenario is to reduce the dark sector 
to one component instead of dark energy and dark matter. This component 
should, at cosmological scales, reproduce both the structure
formation process and the current accelerated expansion of the universe. 
The former condition seems to be the main challenge for such models.

We have compared the ISW signal of UDM models with the $\Lambda$CDM prediction.
Figures (\ref{fullbulk}) and (\ref{cha}) are in agreement with previous 
results, where the background-prefered model parameters of the VDF and the 
GCG imply an unacceptably large amplification of the ISW effect. In fact, 
we confirm and quantify the findings of \cite{barrow} for a wide range of 
parameters.  Although models with $\nu=0$ cannot be ruled out by current data, they 
nevertheless show a significant amplification of the ISW effect that will be detectable 
in the future. 

This tight constraints can be seen as an evidence that either bulk viscous 
effects do not play a role in the cosmic dynamics or that the 
phenomenological ansatz $\xi \sim \rho^{\nu}$ is not appropriate. 
Since the intensive thermodynamic variables are functions of the extensive 
ones, a possible alternative is to describe the bulk viscous pressure in 
terms of energy density and entropy, i.e.~$p=p(\rho,S)$ and 
$\xi=\xi(\rho,S)$. This could imply a well behaved perturbative dynamics 
and alleviate the {\it ISW problem} of such fluids. Recently a microscopic model
for the cosmic bulk viscosity has been introduced as a dark energy candidate in \cite{darkgoo}.
This ``dark goo'' model shows good results when compared with the matter and CMB power spectrum. 
It could be interesting to extend this model to the context of unified dark matter where 
an estimation of the mass of the dark particle can be achieved. An important lesson 
from ``dark goo´´ is that more realistic viscosity coefficients show a complicated dependence on 
the energy density and cannot be written as $\zeta \propto \rho^\nu$. 

On the other hand, we note that bulk viscous pressure represents a small
negative correction to the positive equilibrium pressure. Here we have 
admited the viscous pressure to be the dominating part of the pressure. 
This is clearly beyond the established range of validity of conventional 
non-equilibrium thermodynamics and non-standard interactions are required 
to support such an approach \cite{nonstandard}. Hence, viscous cosmologies 
based on the Israel-Stewart theories \cite{israelstewart} can also be 
considered. Recently, a qualitative analysis of such causal transport 
theory has been performed in \cite{oliver1}.

We have studied the evolution of sub-horizon scales during the matter dominated
epoch. In the standard CDM model the linear growth of small-scale 
perturbations gives rise to dark halos hosting galaxies. Concerning the 
unified scenario, we find a modification of the redshift of 
the matter-radiation equality. As shown in figure \ref{Eq} the prefered 
parameter values for the UDM models are compatible with 
$z^{*}_{\rm eq}>z_{\rm eq}$. Hence, UDM perturbations start to grow earlier
than CDM perturbations. The GCG perturbations follow the CDM
growth $\Delta_{\rm gc}\propto a$ until $z_{nl}$ and consequently, only the
amplitude of the perturbations will be different. On the other hand, 
the VDF perturbations grow in a different way following a nonadiabatic 
Meszaros-like equation derived in section 3. In general, the evolution 
of $\Delta$ is scale-dependent and deviates significantly from $\Lambda$CDM. 
The most important effect at late times is the dominance of 
nonadiabatic contributions causing $\Delta$ to decay rapidly. 

Despite the different evolution of viscous-matter perturbations 
and standard CDM perturbations on subhorizon scales, the
{\it ad hoc} assumption $\delta \xi=0$ can alleviate the $\Delta$ growth
suppression. Since VDF perturbations start to grow before $z_{\rm eq}$,
their amplitudes are of similar size as in the $\Lambda$CDM case. 
In other words, the growth before $z_{\rm nl}$ offsets the late time 
growth suppression. But this assumption is not an acceptable solution and unless other effects like 
shear viscosity or a very different ansatz 
for $\xi$ would lead to qualitatively different results, the VDF models 
are ruled out. This can be interpreted as a complementary probe that 
viscous cosmologies based on
the Eckart formalism (with $\xi \propto \rho^{\nu}$) are strongly challenged as potential contenders for a
general relativity-based description of the cosmic medium.

The inclusion of a baryonic component in the system provides
a more realistic model when compared with the one-fluid approximation 
adopted here. In this case, the background dynamics of the GCG fluid 
remains the same while the VDF will behave differently. However, as 
baryons represent a small fraction 
of the cosmic energy budget we do not expect a significant influence on 
the evolution of the unified dark sector. 

To conclude, the dissipative UDM models considered in this work 
have severe problems to describe 
the observed cosmic structure on largest (enhanced ISW effect) and on 
smallest scales (overdamping due to dissipation). 

\acknowledgments
HV is supported by the CNPq (Brazil) and DAAD (Germany). DJS thanks Deutsche Forschungsgemeinschaft (DFG) for financial support.

\end{document}